\documentclass[11pt,twoside]{article} 
\usepackage{asp2004}
\usepackage{epsf}
\usepackage{psfig}
\usepackage{lscape} 

\begin{document} 
\title{The Structure and Origin of Magnetic Fields on Accreting White Dwarfs}
\author{K.~Reinsch$^1$, F.~Euchner$^1$, K.~Beuermann$^1$, S.~Jordan$^2$,
and B.~T.~G\"ansicke$^3$} 
\affil{$^1$ Univ.-Sternwarte, Geismarlandstr.\,11, 37083 G\"ottingen,
Germany \\
$^2$ Astronomisches Rechen-Institut, 69120 Heidelberg, Germany \\
$^3$ Department of Physics, Univ. of Warwick, Coventry CV4 7AL, UK}

\begin{abstract} 
We have started a systematic study of the field topologies of magnetic single 
and accreting white dwarfs using Zeeman tomography. Here we report on our 
analysis of phase-resolved flux and circular polarization spectra of the 
magnetic cataclysmic variables BL\,Hyi and MR\,Ser obtained with FORS1 at the 
ESO VLT. For both systems we find that the field topologies are more complex 
than a dipole or an offset dipole and require at least multipole expansions
up to order $l = 3$ to adequately describe the observed Zeeman features and
their variations with rotational phase. Overall our model fits are in excellent
agreement with observations. Remaining residuals indicate that the field 
topologies might even be more complex. It is, however, assuring that the 
global characteristics of our solutions are consistent with the average 
effective field strengths and the halo field strengths derived from intensity 
spectra in the past.
\end{abstract}

\section{Introduction}

Magnetism at a detectable level is a common phenomenon among white dwarfs.
According to recent studies the incidence of objects with surface field 
strengths exceeding 2\,MG is at least $\sim$10\% of all single white
dwarfs, and could even be higher (Liebert et al. 2003). A similar fraction 
of $\sim$20\% magnetic systems has been found among accreting white dwarfs in 
cataclysmic variables (G\"ansicke, this volume). Observed field 
strengths range from $\sim$\,1\,kG--1000\,MG with a peak around 16\,MG
in single white dwarfs (Wickramasinghe \& Ferrario 2000, G\"ansicke et 
al. 2002, Schmidt et al. 2003) and 7--230 MG in magnetic CVs
(Beuermann 1998). The frequency distributions of field strengths for both
populations are compared in Fig.\,1.

The origin of the magnetic fields is not well understood. While it is 
reasonable to assume that the white dwarfs with the highest magnetic fields 
evolve from main-sequence Ap and Bp stars, low- and moderate-field magnetic 
white dwarfs appear to imply another origin. 
The decay times of the lowest multipole components are predicted to be long 
compared to the evolutionary ages of the white dwarfs. The magnetic field 
topologies, at least of isolated white dwarfs, are, therefore, likely to be 
relics of previous evolutionary phases. In accreting systems, the field 
structure in the outer layers of the white dwarf, however, may have been 
significantly changed if the accretion rate is high enough that accretion 
occurs more rapidly than ohmic diffusion (Cumming 2002) and Rayleigh-Taylor 
instabilities (Romani 1990).

\begin{figure}[!ht]
\plotone{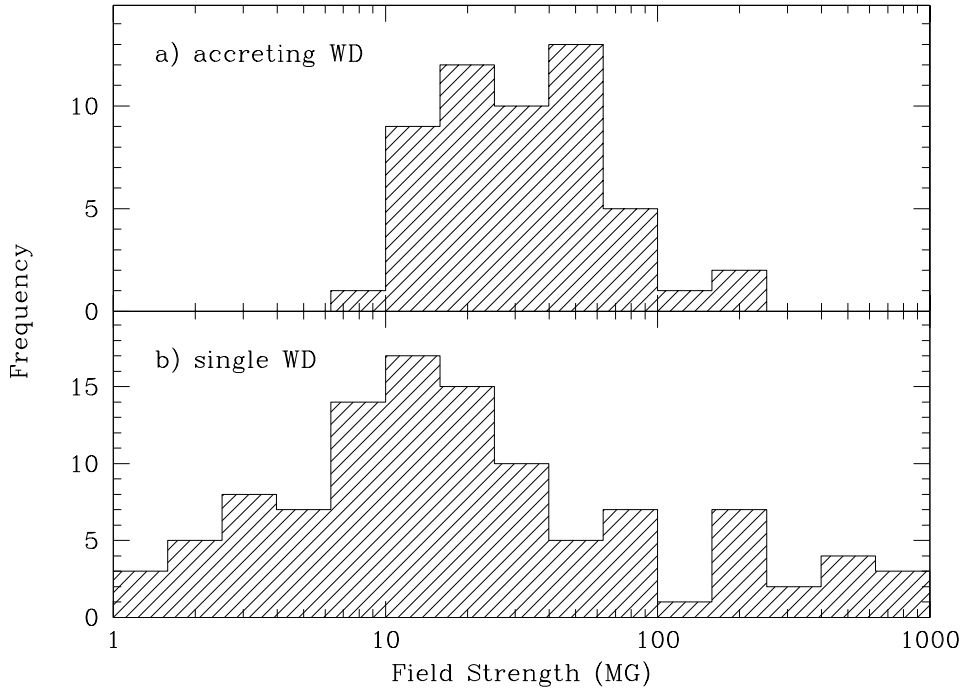}
\caption{Frequency distributions of the field strengths of accreting 
{\it (top)} and single {\it (bottom)} white dwarfs. For the former, average
field strengths in the main accretion region have been used of all 53 polars 
for which reliable measurements are available (cf. Beuermann 1998). The 
statistics for single white dwarfs include data from the compilation  
by Wickramasinghe \& Ferrario (2000) and  white dwarfs from the SDSS 
(Schmidt et al. 2003). The distribution for magnetic CVs is biased by 
the omission of low-field systems (intermediate polars) for which only 
field strength estimates exist and probably also by selection effects 
which hamper the discovery of high-field polars.
}
\end{figure}

\section{Zeeman tomography}

Zeeman tomography is a systematic method which we have developed to derive the 
surface magnetic field structure on rotating white dwarfs (Euchner et al.
2002). It utilizes a least-squares optimization code based on an evolutionary 
strategy to reconstruct the multipole parameters of the field from a series of 
flux and circular polarization spectra obtained at different rotational phases. 
The observational data are fitted with a database of precomputed Zeeman spectra 
of homogeneous magnetic white dwarf atmospheres calculated with a code 
developed by S. Jordan for a grid of field strengths $B$, effective 
temperatures $T$, and angles $\psi$ between the field direction and the line of 
sight. 

\section{Observations}

\begin{figure}[!ht]
\plotone{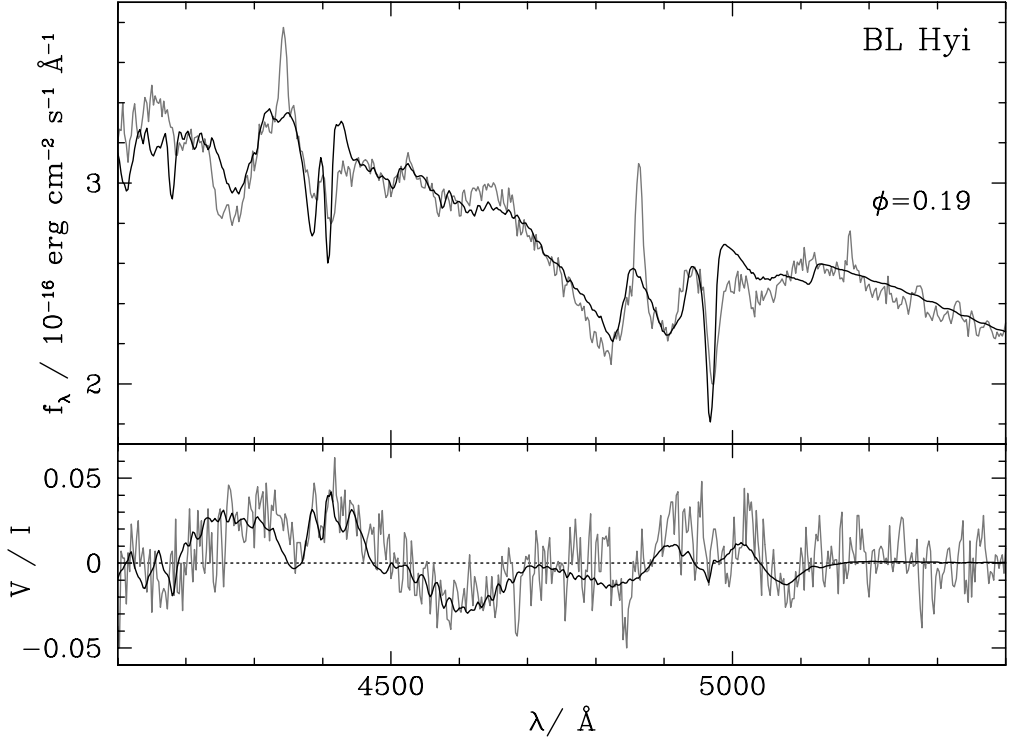}
\caption{Flux {\it (top)} and circular polarization {\it (bottom)} spectra of 
BL\,Hyi at rotational phase $\Phi = 0.19$. 
The synthetic spectra for the best-fit model {\it (black line)} consisting of 
a truncated multipole expansion up to order $l = 3$ are shown superimposed on 
the observed spectra {\it (grey curve)}.}
\end{figure}

\begin{figure}[!ht]
\plottwo{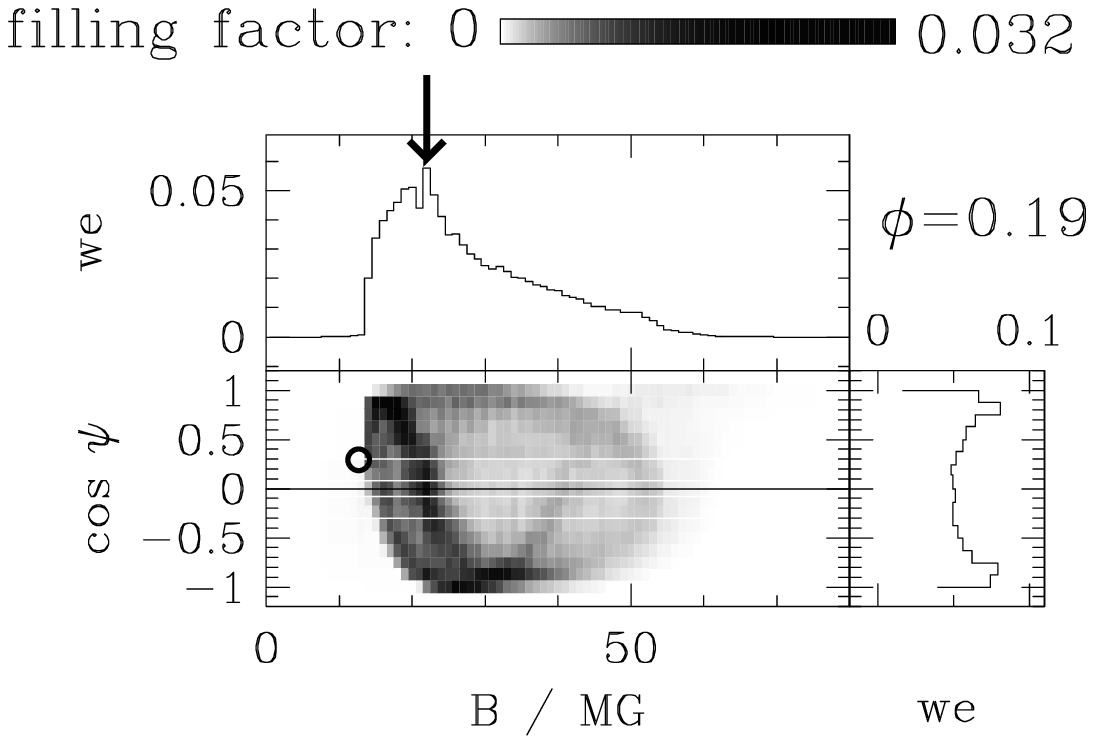}{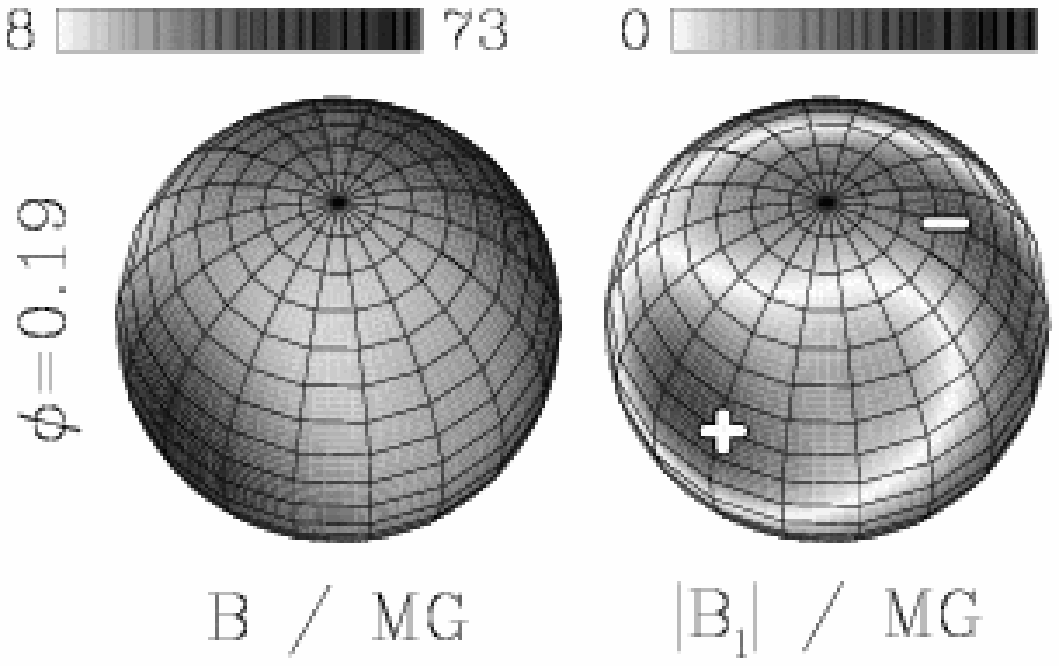}
\caption{{\it Left:} Frequency distribution of the magnetic field strength 
$|\vec{B}|$ and the viewing direction cosine of $\vec{B}$, $\cos \psi$. 
{\it Right:} Distribution of $|\vec{B}|$ and its longitudinal component 
$B_{\rm l}$ on the white dwarf in BL\,Hyi at rotational phase $\Phi = 0.19$.
The vertical arrow in the left diagram marks the average effective photospheric
field strength, $B_{\rm eff} = 22$\,MG; the open circle depicts the halo field
strength, $B_{\rm halo} = 12$\,MG (Schwope et al. 1995) with the location
of the main accretion pole derived by Piirola et al. (1987).}
\end{figure}

\begin{figure}[!ht]
\plotone{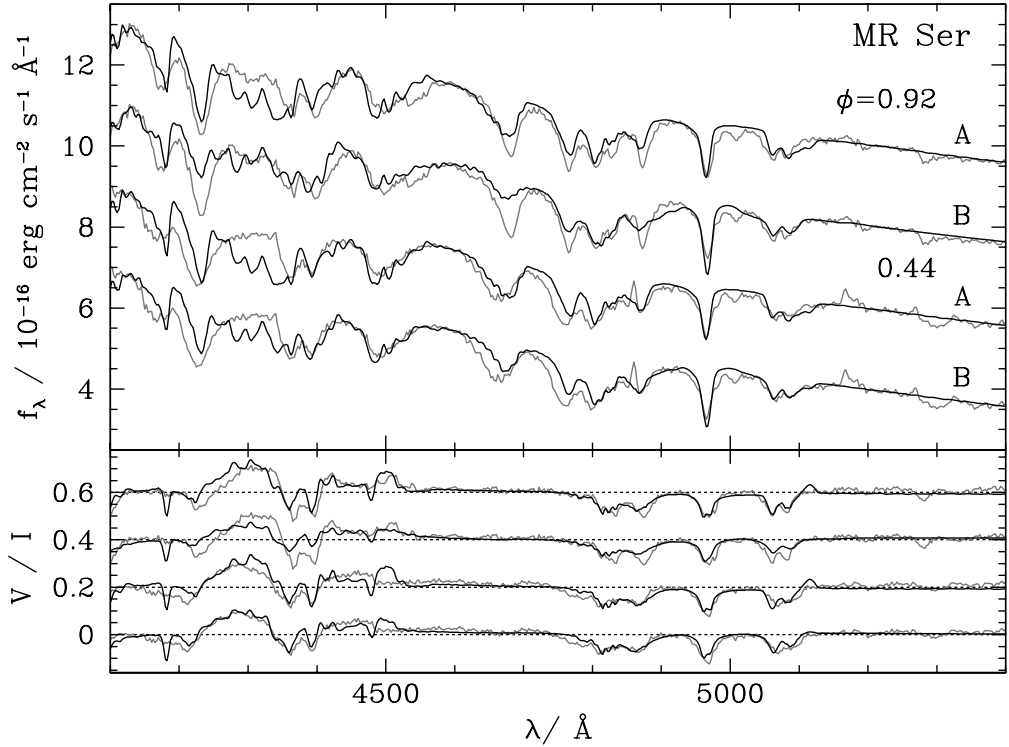}
\caption{Flux {\it (top)} and circular polarization {\it (bottom)} spectra of 
MR\,Ser at $\Phi =$ 0.92 (faint phase) and $\Phi =$ 0.44 (bright phase). For 
each phase the observed spectra {\it (grey curves)} are compared with two 
different models {\it (black lines)}: {\it (A)} an offset dipole field, {\it 
(B)} a multipole expansion up to order $l = 3$.
For clarity, the upper three spectra have been offset by 2, 4, and 6 flux 
units, and the polarization spectra by 0.2, 0.4, and 0.6 units, respectively.
}
\end{figure}

Within our framework of systematic field topology studies of single and 
accreting magnetic white dwarfs we have obtained spin-phase resolved circular 
spectropolarimetry of the magnetic cataclysmic variables BL\,Hyi and 
MR\,Ser. Data were taken in multi-object spectropolarimetric mode (PMOS) with 
FORS1 at the ESO VLT while both systems were in low states of accretion. 

During data reduction instrumental polarization effects and linear polariza\-tion 
cross-talk have been taken into account by computing circular polarization
($V/I$) spectra from two subsequent exposures taken with retarder plate 
position angles $\phi = -45\deg$ and $\phi = +45\deg$. Atmospheric 
absorption losses have been corrected for using the spectra of simultaneously
observed field stars. Wavelength-dependent flux losses due to varying seeing
conditions have been compensated for by normalizing the flux spectra to a mean
continuum level.

\section{Results}

\subsection{BL\,Hyi}

Spectropolarimetric observations of BL\,Hyi have been obtained on 4 December 
1999 and cover one complete rotational period of the accreting white dwarf. 
From the data we have extracted flux and polarization spectra at five 
rotational phases according to the ephemeris of Wolff et al. (1999). 
In Fig.\,2 we show a comparison of our observations and the best fit model
near the maximum of the bright phase ($\Phi = 0.19$), i.e. when the main 
accretion region is visible.
The observed Zeeman features of the Balmer lines H$\alpha$, H$\beta$, and 
H$\gamma$ vary significantly with rotational phase and require a multipole 
expansion up to order $l = 3$ to describe the field topology (Fig.\,3).

Our best fit solution reveals a complex field topology but is consistent with
the main characteristics like the average effective photospheric field 
strength, $B_{\rm eff} = 22$\,MG, and the halo field strength, $B_{\rm halo} 
= 12$\,MG, derived from intensity spectra alone (Schwope et al. 1995).

\subsection{MR\,Ser}

On 23 May and 24 May 2004 we have obtained 5.7 hours of spectropolarimetry of 
MR\,Ser equivalent to three rotational periods. The extracted flux and 
polarization spectra have been grouped into six rotational phase bins 
according to the ephemeris given by Schwope et al. (1993).
In Fig.\,4 we show the faint phase ($\Phi = 0.92$) and the bright phase 
($\Phi = 0.44$) spectra, respectively. We have obtained two possible solutions 
for the field topology (Figs.\,5 and 6). The first one consists of a dipole 
with $B = 53$\,MG which has been offset along the dipole axis by 0.19 white 
dwarf radii. The second model is a truncated 
multipole expansion up to order $l = 3$ and provides a slightly better fit 
to the data. Again, our more complex field topologies match well the global 
field characteristics ($B_{\rm eff} = 28$\,MG, $B_{\rm halo} = 24$\,MG) 
observed previously (Schwope et al. 1993).
\begin{figure}[!ht]
\plottwo{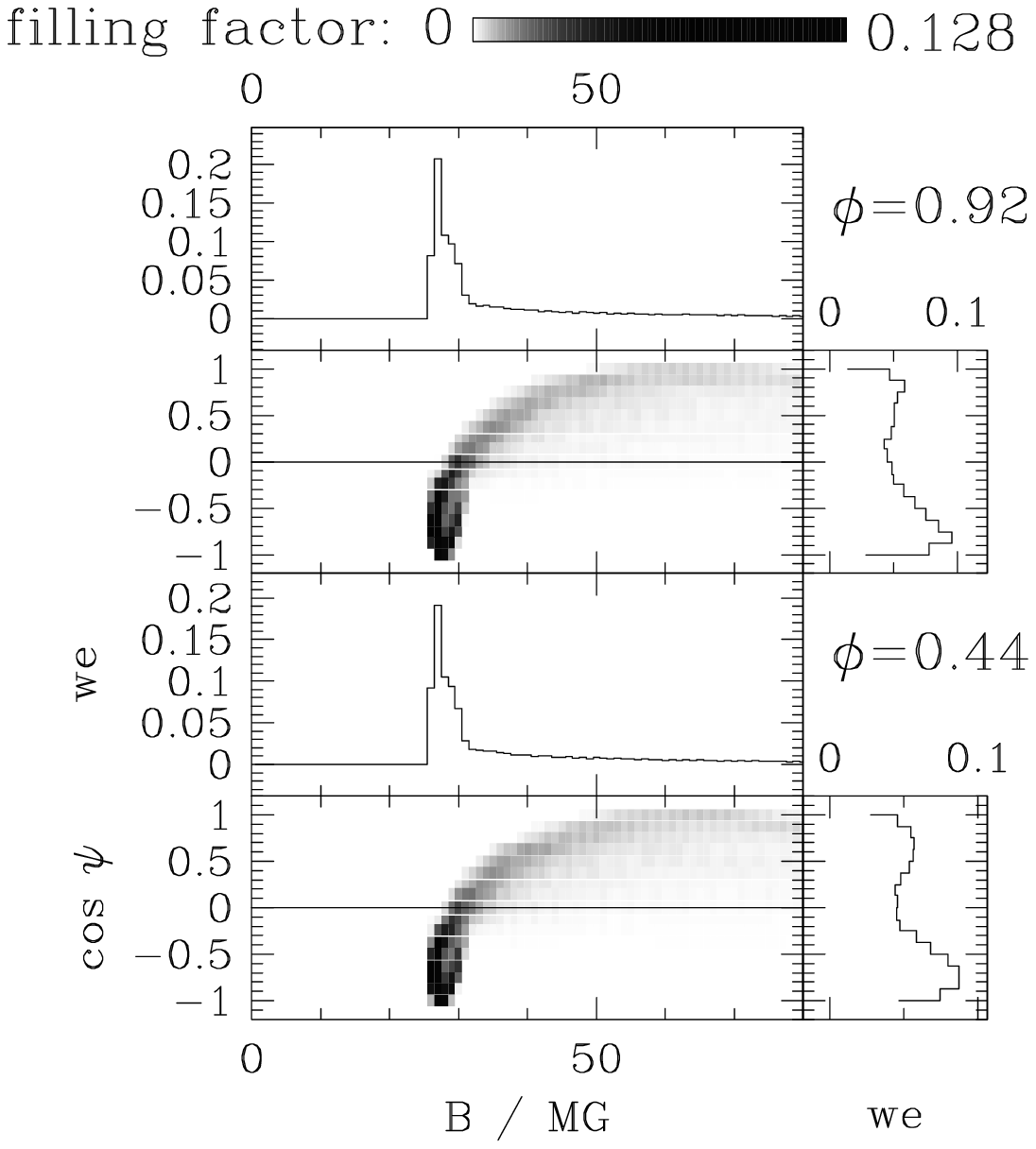}{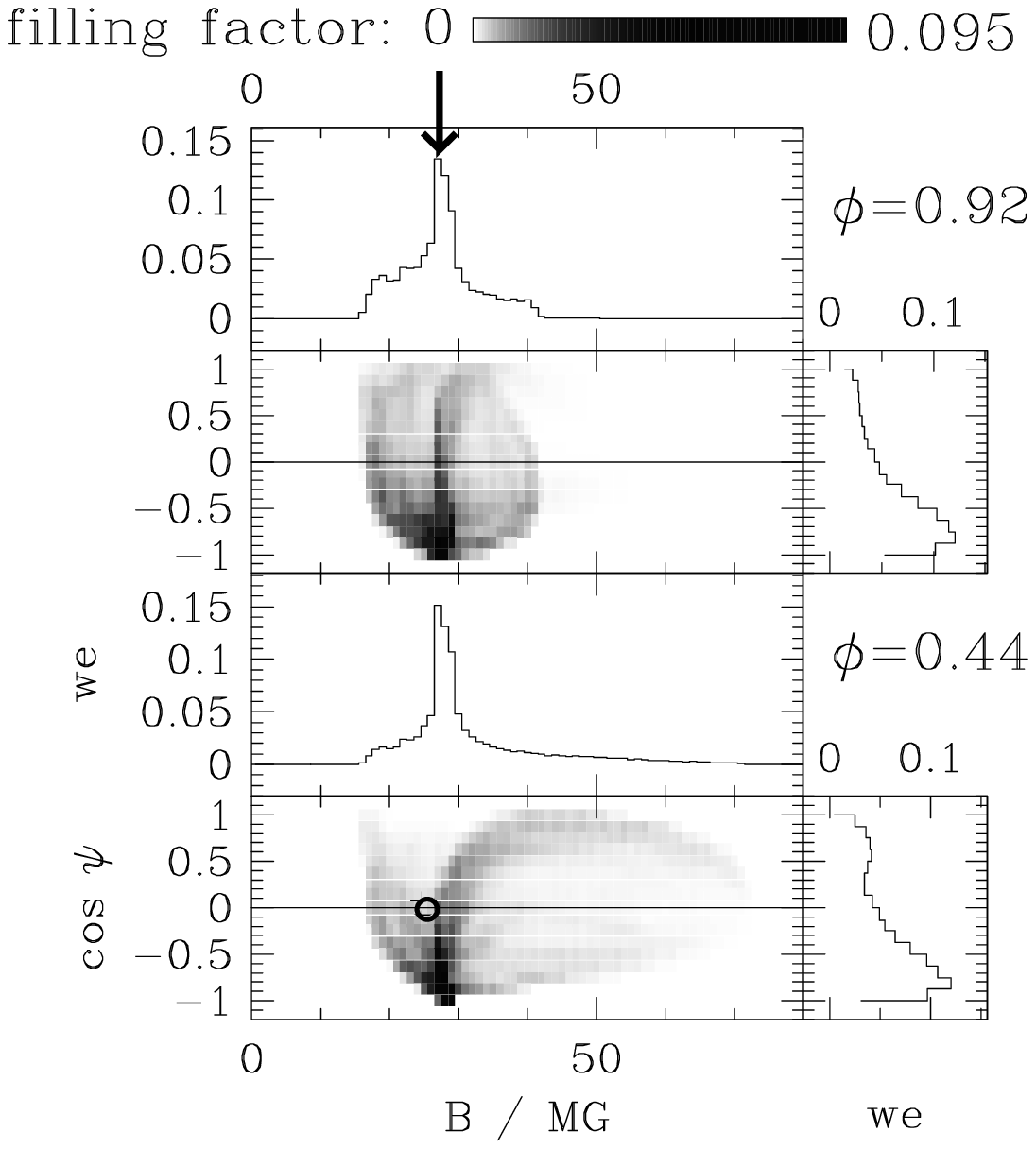}
\caption{Frequency distribution of the magnetic field strength $|\vec{B}|$ and 
the viewing direction cosine of $\vec{B}$, $\cos \psi$, on the white dwarf in 
MR\,Ser at rotational phases $\Phi =$ 0.92 and $\Phi =$ 0.44, respectively, 
derived for the two models: {\it (left)} offset dipole, {\it (right)} truncated 
multipole.
The vertical arrow in the right diagram marks the average effective photospheric
field strength, $B_{\rm eff} = 28$\,MG; the open circle depicts the halo field
strength, $B_{\rm halo} = 24$\,MG, reported by Schwope et al. (1993).}
\end{figure}

\begin{figure}[!ht]
\plottwo{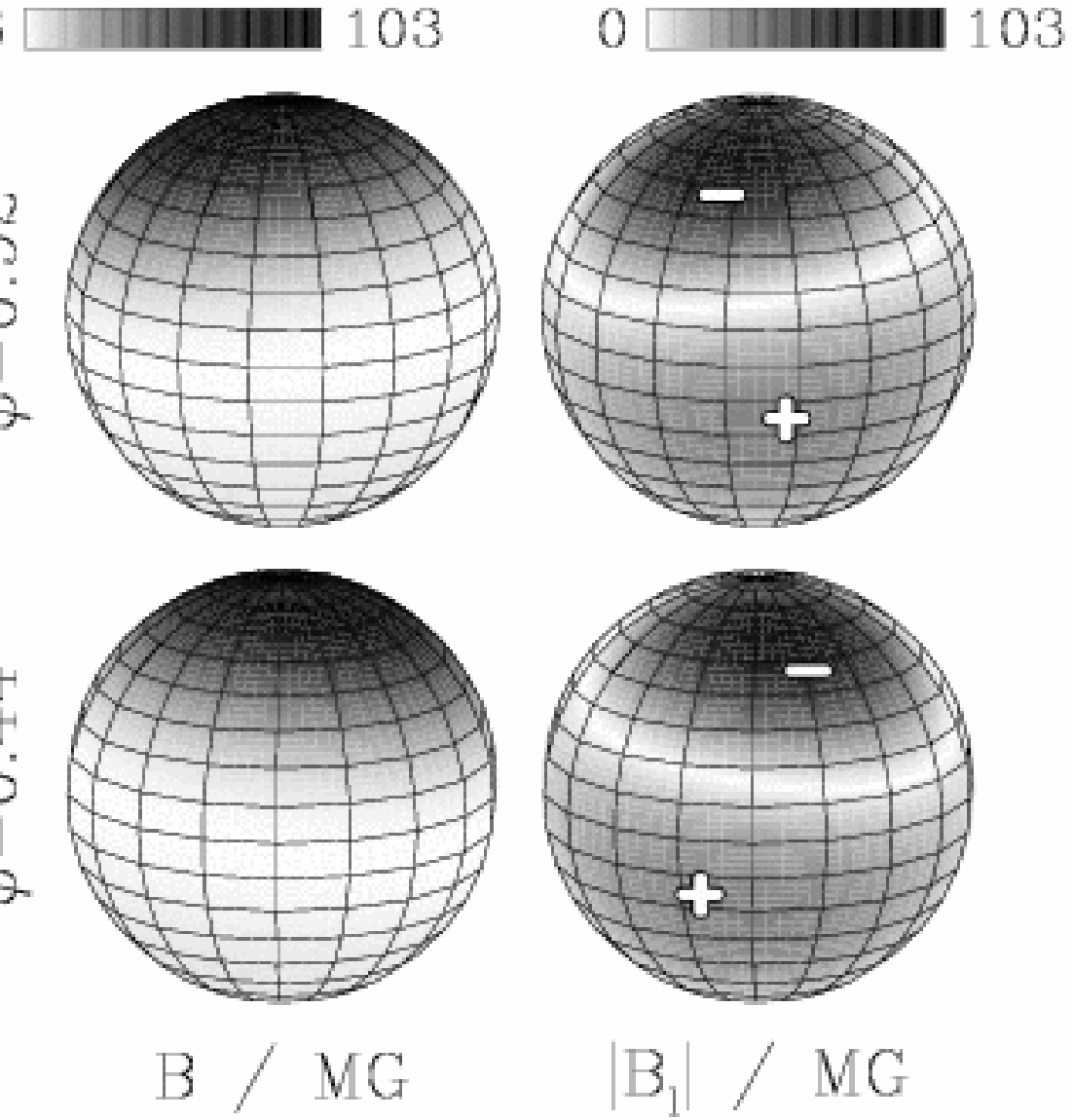}{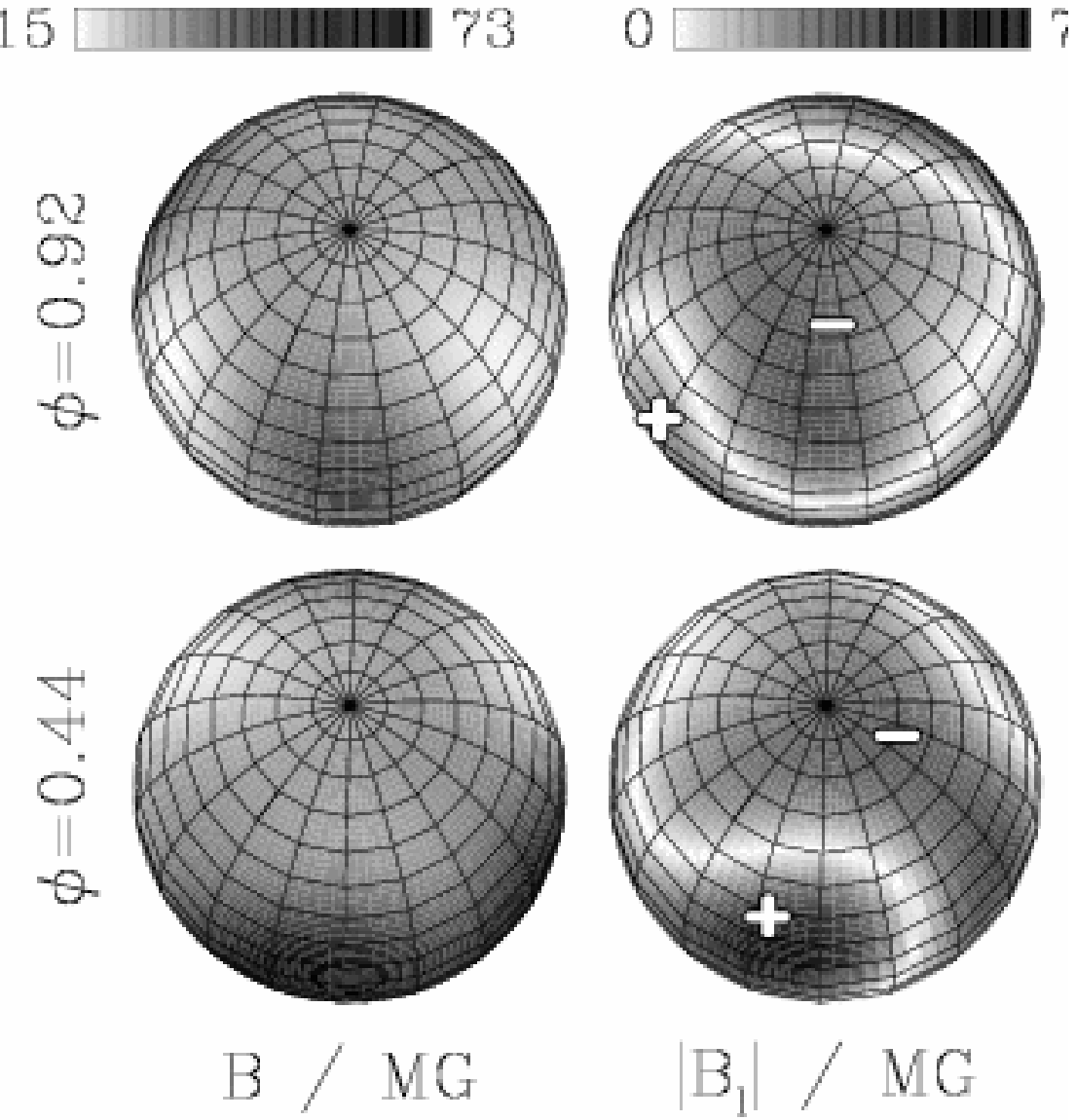}
\caption{Distribution of the total magnetic field strength $B$ and its 
longitudinal component $B_{\rm l}$ on the white dwarf in MR\,Ser at 
rotational phases $\Phi =$ 0.92 and $\Phi =$ 0.44, respectively, derived
for the two models: {\it (left)} offset dipole, {\it (right)} truncated 
multipole.}
\end{figure}

\section{Conclusions}

The Zeeman tomographical analysis of phase-resolved spectropolarimetry 
provides for the first time detailed information about the range of
field strengths and the field topology of accreting white dwarfs.  
For both systems discussed here we find that at least multipole expansions 
up to order $l = 3$ are required to describe the field topologies.
Remaining residuals indicate that the field topologies might even be more 
complex.

\acknowledgements{Based on observations collected at the European Southern 
Observatory, Chile under program numbers 64.P-0150(C) and 073.D-0322(B).}

\end{document}